\documentclass[prl,showpacs,preprintnumbers,amsmath,amssymb,endfloats]{revtex4}

\usepackage{graphicx}
\usepackage{dcolumn}
\usepackage{bm}

\begin{document}


\title{Field-induced spin mixing in ultrathin superconducting Al and Be
    films in high parallel magnetic fields}

\author{P.W. Adams}
\affiliation{Department of Physics and Astronomy\\Louisiana State University\\Baton Rouge, Louisiana,
70803}%


\date{\today}

\begin{abstract}
We report spin-dependent electron density of states (DOS) studies of ultra-thin superconducting Al
and Be films in high parallel magnetic fields.  Superconductor-insulator-superconductor (SIS) tunneling spectra are
presented in which both the film and the counterelectrode are in the paramagnetic limit.  This SIS configuration is
exquisitely sensitive to spin mixing and/or spin flip processes which are manifest as DOS singularities at
$eV=2\Delta_o\pm eV_z$.  Both our Al and Be data show a well defined subgap peak whose magnitude grows dramatically
as the parallel critical field is approached.  Though this feature has previously been attributed to spin-orbit
scattering, it is more consistent with fluctuations into a field induced mixed-spin state.   
\end{abstract}

\maketitle

	With recent discoveries of itinerant ferromagnetic superconductors,
represented by  UGe$_2$ \cite{UGe2} and ZrZn$_2$ \cite{ZrZn2}, and Fulde-Ferrell-Larkin-Ovchinnikov (FFLO)
superconductivity in CeCoIn$_5$ \cite{FFLO,CeCoIn}, research on systems exhibiting a non-trivial interplay between
magnetism and superconductivity has moved to the forefront on condensed matter physics.  In this Letter we probe the spin
states of superconducting Al and Be films in high parallel magnetic fields via spin polarized electron tunneling
measurements \cite{Giaever}.  The films are sufficiently thin so as to restrict the transverse motion of electrons, thus
allowing us to access the high field regime while maintaining time reversal symmetry
\cite{AleinerPRL} up to the Clogston-Chandrasekhar critical field
$H_{c||}=\sqrt{2}\Delta_o/(g\mu_B)$, where g is the Land\'{e} g-factor, $\mu_B$ is the Bohr magneton, and
$\Delta_o$ is the superconducting gap \cite{Clogston}. Though the films are too disordered to support a FFLO phase
\cite{Fulde}, they are a model system for studying the spin states of BCS superconductivity in the presence of a
non-negligible Zeeman field that ultimately drives the first-order spin-paramagnetic transition associated with
$H_{c||}$ \cite{Fulde, AdamsAl} and the long conjectured FFLO regime just above $H_{c||}$ \cite{Fulde}. 
Tunneling measurements in fields $H_{||}\gtrsim\frac{1}{2}H_{c||}$ reveal a subgap peak in the DOS spectrum, shifted down
from the primary BCS peak by the Zeeman energy.  The magnitude of the satellite peak varies as the square root of the
reduced field.  Though this peak has previously been attributed to spin-orbit (SO) scattering in Al \cite{MeserveyMixing},
it is also manifest in the much lighter element Be, suggesting that it is a property of the high field condensate.

In the mid 1970's Tedrow, Meservey, and coworkers conducted a series of tunneling experiments on paramagnetically
limited Al films.  They showed that the tunneling spectrum of a superconductor-insulator-superconductor (S-I-S) junction,
in which both the film and the counter-electrode are thin, will not exhibit a Zeeman splitting so long as there is
no spin mixing or spin flip processes \cite{Fulde}.  Assuming that the gap is $\Delta_o$ on either side of the
junction, then the tunneling spectrum has a single BCS peak at the usual $|eV|=2\Delta_o$, independent of the
Zeeman energy $eV_z=g\mu_BH_{||}$, where $e$ is the electron charge.  If, however, there is spin flip during the
tunneling, then satellite peaks will appear at energies $|eV|=2\Delta_o\pm eV_z$ \cite{Fulde,MeserveyMixing}. 
Similarly, if there is a mechanism by which the spin eigenstates are partially mixed, then there will be a
minority-spin satellite peak in the spectrum at $|eV|=2\Delta_o - eV_z$ \cite{MeserveyMixing}.  No spin flip
effects have ever been reported for tunneling through standard non-magnetic oxides such as Al$_2$O$_3$.  However,
spin mixing has been observed, the origin of which is the primary focus of this Letter.

	Spin-orbit scattering is known to cause spin mixing and in thin films the SO scattering rate,
$1/\tau_{so}$, increases with increasing atomic mass $Z$ as
$\tau_{s}/\tau_{so}\sim Z^4$, where $\tau_{s}$ is the surface scattering time \cite{MeserveySO}, suggesting that
light elemental films are the best candidates for purely spin singlet superconductivity.  Ironically, the first
direct electron tunneling evidence of spin mixing was obtained in thin Al films using the S-I-S
configuration described above \cite{MeserveyMixing}.  A small subgap peak in the tunneling DOS
was seen in 5 nm thick crossed Al films at the voltage $|eV|=\Delta_{Film1}+\Delta_{Film2}-eV_z$, consistent with a
finite SO scattering rate.  However, earlier measurements of the Knight shift in Al films showed that the shift
extrapolated to zero at T=0 in accord with BSC theory, suggesting that $1/\tau_{so}\sim0$ in Al \cite{Knight}.  More
recent studies of the spin paramagnetic transition in Al and Be films have revealed both tricrtical point
behavior \cite{AlTCP,BeTCP} and quasi-coherent fluctuation modes \cite{FBTA} that are inconsistent with a
finite SO scattering rate \cite{Fulde,FBTA}.   

 The Al (Be) films used in these experiments were made by e-beam deposition of 3 - 5 nm of $99.999\%$ Al ($99.5\%$
Be) onto fire polished glass microscope slides cooled to 84 K.  Typical deposition rates were $\sim0.1$ nm/s in a
vacuum of $\sim0.5$ $\mu$Torr.  The Al (Be) films had a transition temperature $T_c\sim2.7$ K ($T_c\sim0.5$ K) and a
parallel critical field $H_{c||}\approx6.0$ T ($H_{c||}\approx1.0$ T).  Tunnel junctions were formed by exposing
the films to atmosphere for $0.2 - 1$ hours in order to form a native oxide.  Then an Al (Be) counter-electrode of
the same thickness as the film was deposited directly on top of the film at 84 K.  The integrity of the junctions was
tested by measuring the dc I-V characteristics in zero magnetic field at T=50 mK.  All of the tunneling data presented
below are S-I-S.  The films were aligned to within $0.1^o$ of parallel by an {\it in situ} mechanical rotator. 

	In Fig.\ 1 we plot the zero field tunneling conductance as a function of bias voltage for a 2.7-nm Al film and a
4-nm Be film, each with a normal state sheet resistance $R\sim 1$ k$\Omega$.  In order to compare the Al
and Be spectra, we have normalized the bias voltage by each film's respective gap.  Within the resolution of the
measurements, it is reasonable to assume that the superconducting gap of a film is the same as that of its
counter-electrode $\Delta_{Film}\sim\Delta_{CE}\sim\Delta_{o}$, where $\Delta_o$ is the zero temperature, zero field
gap.  At low temperatures the tunneling conductance is directly proportional to the quasiparticle density of states
\cite{Tinkham}.  Note the very sharp BCS DOS peaks at $2\Delta_o$ in Fig.\ 1.  In Fig.\ 2 we show the spectra at
several subcritical values of $H_{||}$.  The Al spectra are shown in a semi-log plot in order to better display the
subgap features, and the Be curves have been shifted for clarity.  Note that the position of the primary peak is
relatively insensitive to field even at fields very near $H_{c||}$, and that, as expected, it displays no Zeeman
splitting.  In contrast, there is a subgap feature whose magnitude and position are a function of field.  The Al data
in Fig.\ 2, displays the spin mixing feature first reported in Al by Meservey and Tedrow, though the peaks
are somewhat sharper than those in Ref.\ 10 due to fact that the data were taken at 50 mK as opposed to 400 mK. 
Assuming that the surface scattering rates of the Al and Be films are comparable, then $1/\tau_{so}$ of the Be film
should be two orders of magnitude smaller than that of Al.  Consequently, it seems unlikely that the supgap peaks
in the Be data are due to SO scattering, which calls into question its role in the Al data.  (The zero bias peaks in the Be
spectra are a finite temperature effect, see Fig.\ 4 caption).  Broken inversion
symmetry can also produce a mixing of the spin singlet and triplet pairings in the presence of SO scattering
\cite{Inversion}.  However, this mechanism appears to require an intrinsic SO scattering rate that is
inconsistent with the tricritical point behavior of the films.

	The positions of the primary and subgap peaks, such as those in Fig.\ 2, are plotted in Fig.\ 3.  For data close to the
critical field, we were careful to insure that both the film and the counter-electrode were superconducting by monitoring
the in-plane resistivity of each.  We have normalized the voltage and field axes by the gap values in order to collapse
the data sets.  The weak quadratic field dependence of primary peak position, shown as solid symbols, is due to pair
breaking \cite{Tinkham}.  The open symbols represent the position of the subgap features.  Both the Al and Be data fall on
the dashed line which represents the zero field gap minus the Zeeman energy assuming g=2.  Indeed, there is sufficient
resolution in the peak positions to rule out the possibility of a weak S-I-N component of the spectrum which would produce
peaks at $|eV|=\Delta_o\pm eV_z/2$.

 To gain a better understanding of the origin of the subgap peak, we studied its dependence on temperature and field
orientation.  By rotating slightly out of parallel orientation in a sub-critical field, we could test the
effect of breaking time reversal symmetry.  If the subgap peaks arise from a quasi-coherent fluctuation mode, then they
will likely require this symmetry \cite{FBTA}.  This is indeed the case as can be seen in the top panel of Fig.\ 4, where
a misalignment of only $2^o$ completely washes out the feature, though the position of the primary peaks
remains unchanged.  In the bottom panel of Fig.\ 4 we compare a Be spectrum at 50 mK and 200 mK.  The central peak in the
200 mK curve is the well known S-I-S finite temperature peak which occurs at
$|\Delta_{Film}-\Delta_{CE}|\sim0$ \cite{Tinkham}.  Note that the Zeeman subgap peak is somewhat attenuated at higher
temperature, and is therefore not activated.

	It is particularly evident that the magnitude of the subgap peaks in the spectra of Fig.\ 2 grows as one approaches
the spin-paramagnetic transition.  We have subtracted the field dependent background from the peaks to get a
relative measure of the peak magnitudes.  Figure 5 shows the subgap peak magnitude as a function of the square-root
of the reduced field $[1-H_{||}/H_{c||}]^{-1/2}$.  Though this scaling form was chosen empirically, the linearity
of the data strongly suggests critical behavior associated with a fluctuation mode which is being stabilized by the
parallel field.  

	In conclusion we believe that the spin-mixing feature is an intrinsic
property of spin-paramagnetically limited BCS superconductivity and that the ground state of the system is significantly
altered near the critical field.  It seems likely that the observed mixing has implications for FFLO physics to the
extent that the films are believed to have a stable FFLO phase just above $H_{c||}$ in the zero scattering limit. 
Clearly, in the absence of disorder the system must find a way to evolve from spin singlet zero momentum pairing to the
finite momentum depaired state of FFLO \cite{FFLO2}.  The nature of this process, particularly in the presence of disorder,
remains an open question.      

	We gratefully acknowledge enlightening discussions with Robert Meservey, Kun Yang, Shivaji Sondhi, Igor Aleiner,
Dana Browne, R.G. Goodrich, and David Young. This work was supported by the National Science Foundation under Grant
DMR 02-04871.

\newpage
\begin{figure}
\includegraphics[width=3.5in]{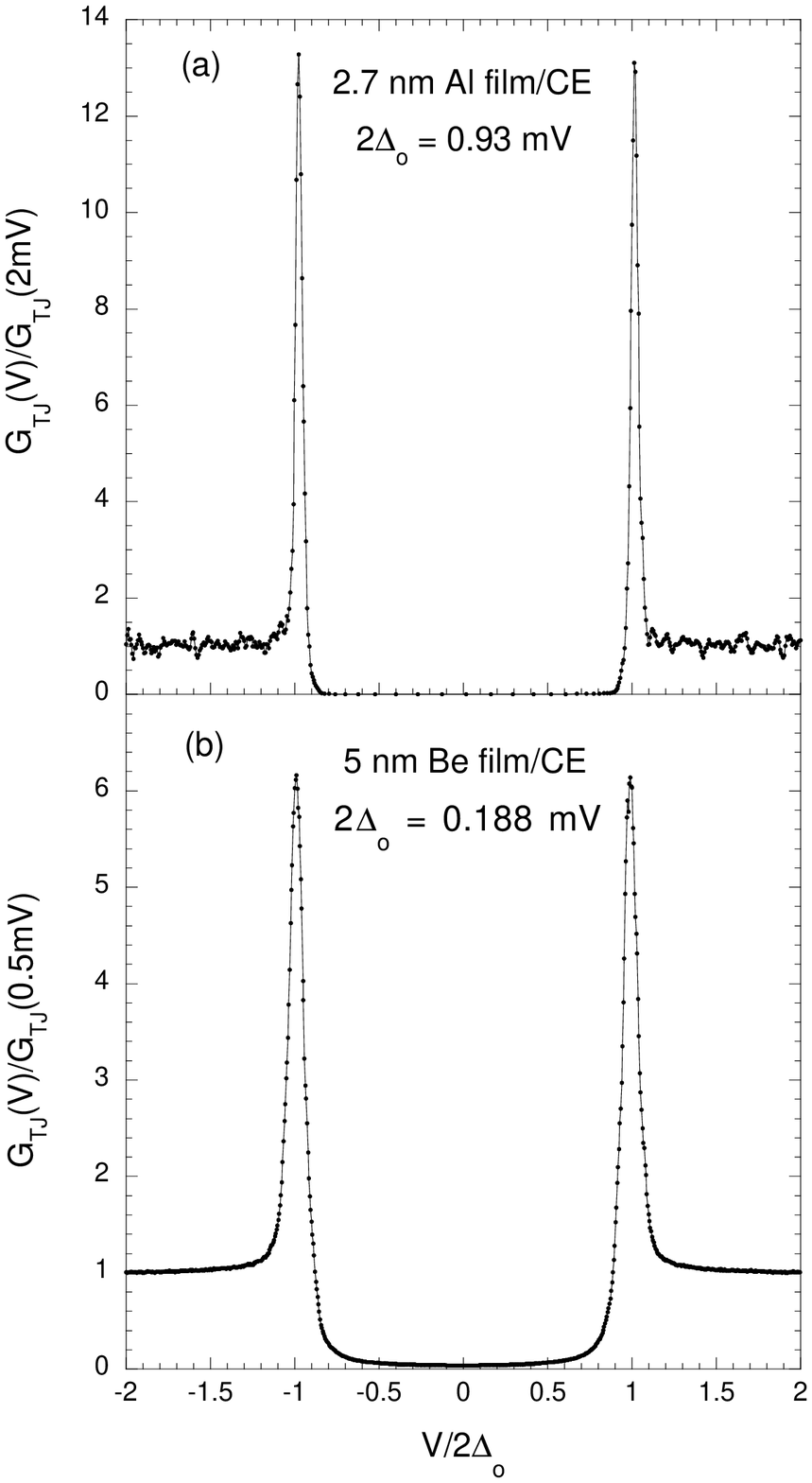}
\caption{\label{fig:epsart} S-I-S tunneling in zero field at T=50 mK for an Al film (A) and Be film (B).  The films
and their respective counter-electrodes were identical.  The tunnel junction conductances are plotted as a function
of the bias voltage normalized by the sum of the zero field gaps $2\Delta_o=\Delta_{Film}+\Delta_{CE}$.}
\newpage
\end{figure}

\begin{figure}
\includegraphics[width=3.5in]{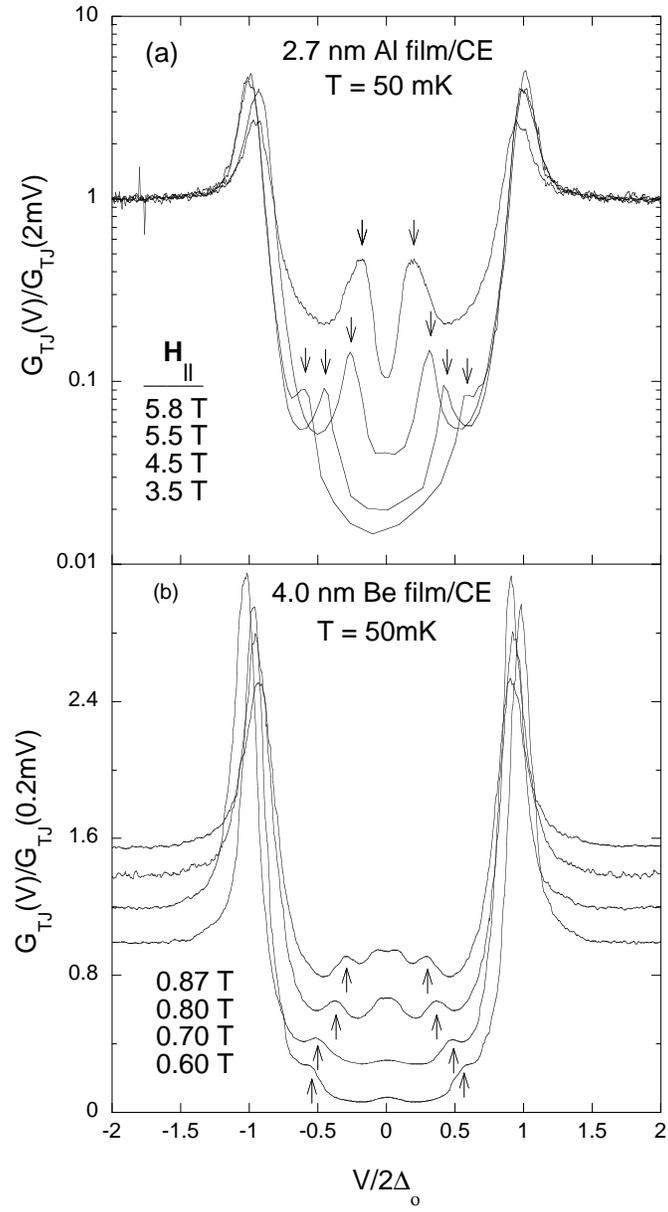}
\caption{\label{fig:epsart}  Tunneling spectra of the films in Fig.\ 1 at several values of parallel magnetic
field, where $H_{c||}=$ 5.9 T for the Al film, and 0.9 T for the Be film.   The arrows show the location of
the subgap peaks.  The Be curves have been shifted for clarity.}
\newpage
\end{figure}

\begin{figure}
\includegraphics[width=5in]{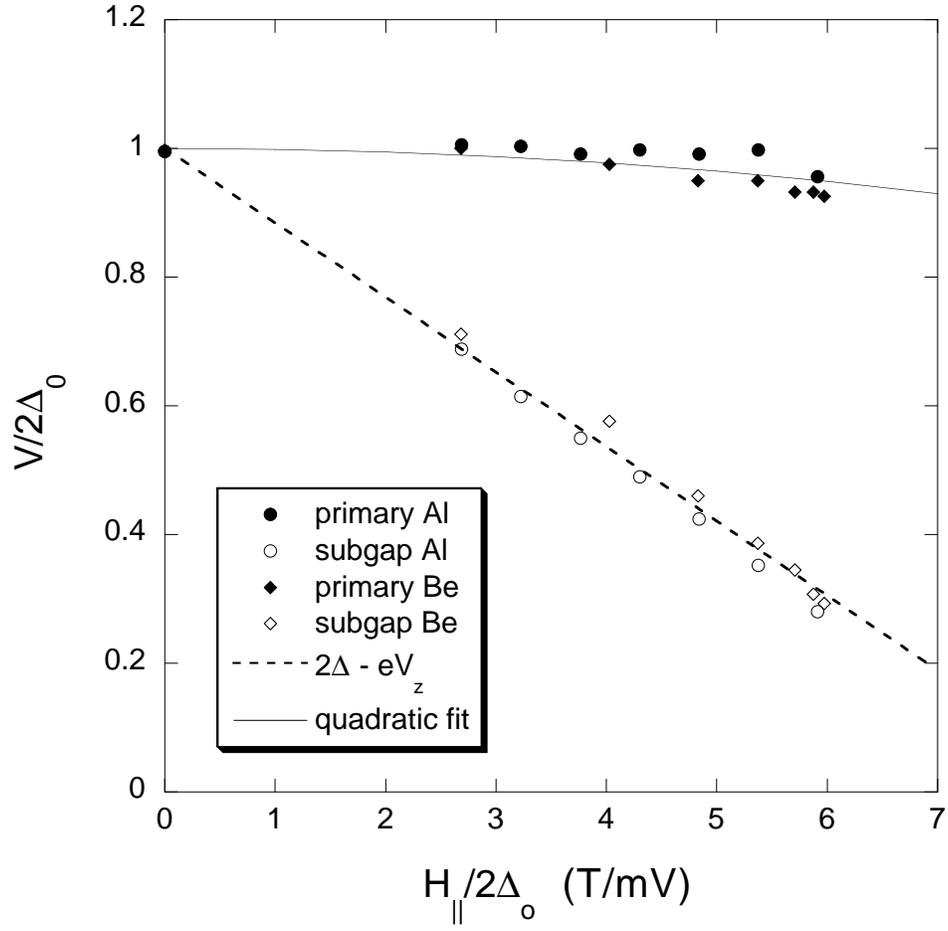}
\caption{\label{fig:epsart} Voltage positions of the primary and subgap peaks from spectra such as that in Fig.\ 2. 
The voltages and fields have been normalized by $2\Delta_o$ in order to collapse the data sets. 
The dashed line represents $2\Delta_o-eV_z$, where $eV_z$ is the Zeeman energy.}
\newpage
\end{figure}

\begin{figure}
\includegraphics[width=3.5in]{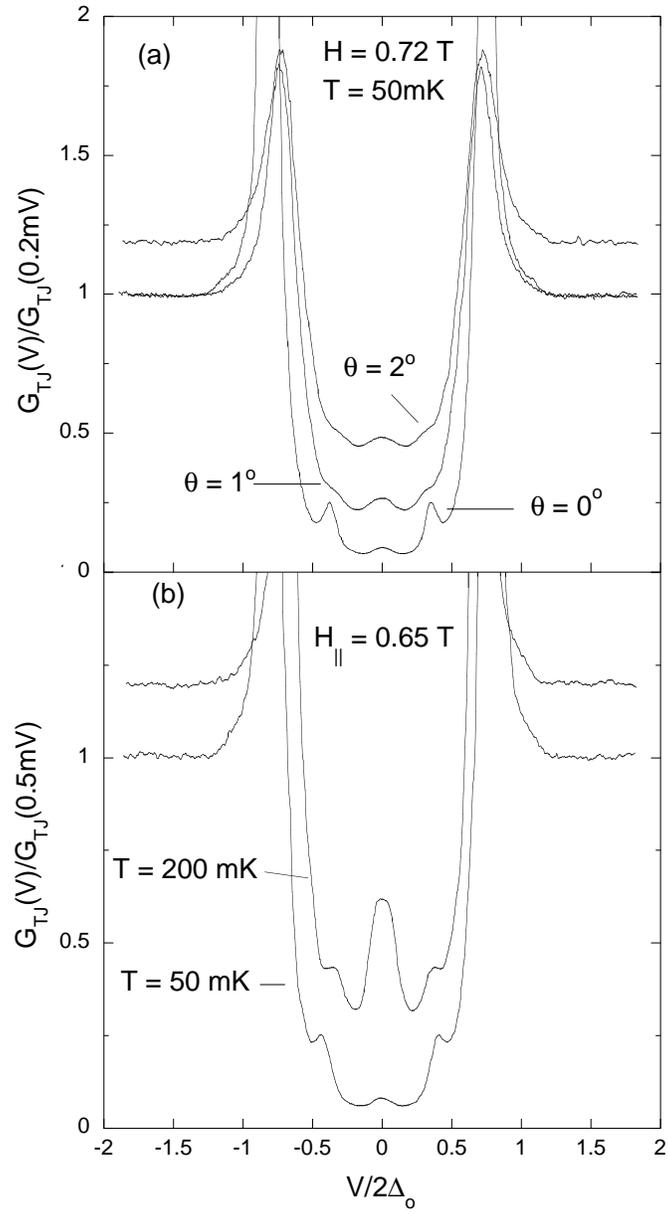}
\caption{\label{fig:epsart} (A) Effect of rotation out of parallel orientation on the tunneling spectra of a 4 nm Be
film.  Note the sensitivity of the subgap peak to $\theta$, where $\theta=0^o$ corresponds to parallel orientation. 
The $\theta=2^o$ curve has been shifted for clarity.  (B) Tunneling spectra of the Be film at two different
temperatures.  The central peak in the 200 mK curve is the usual S-I-S finite temperature peak at
$|\Delta_{Film}-\Delta_{CE}|\sim0$.}
\newpage
\end{figure}

\begin{figure}
\includegraphics[width=5in]{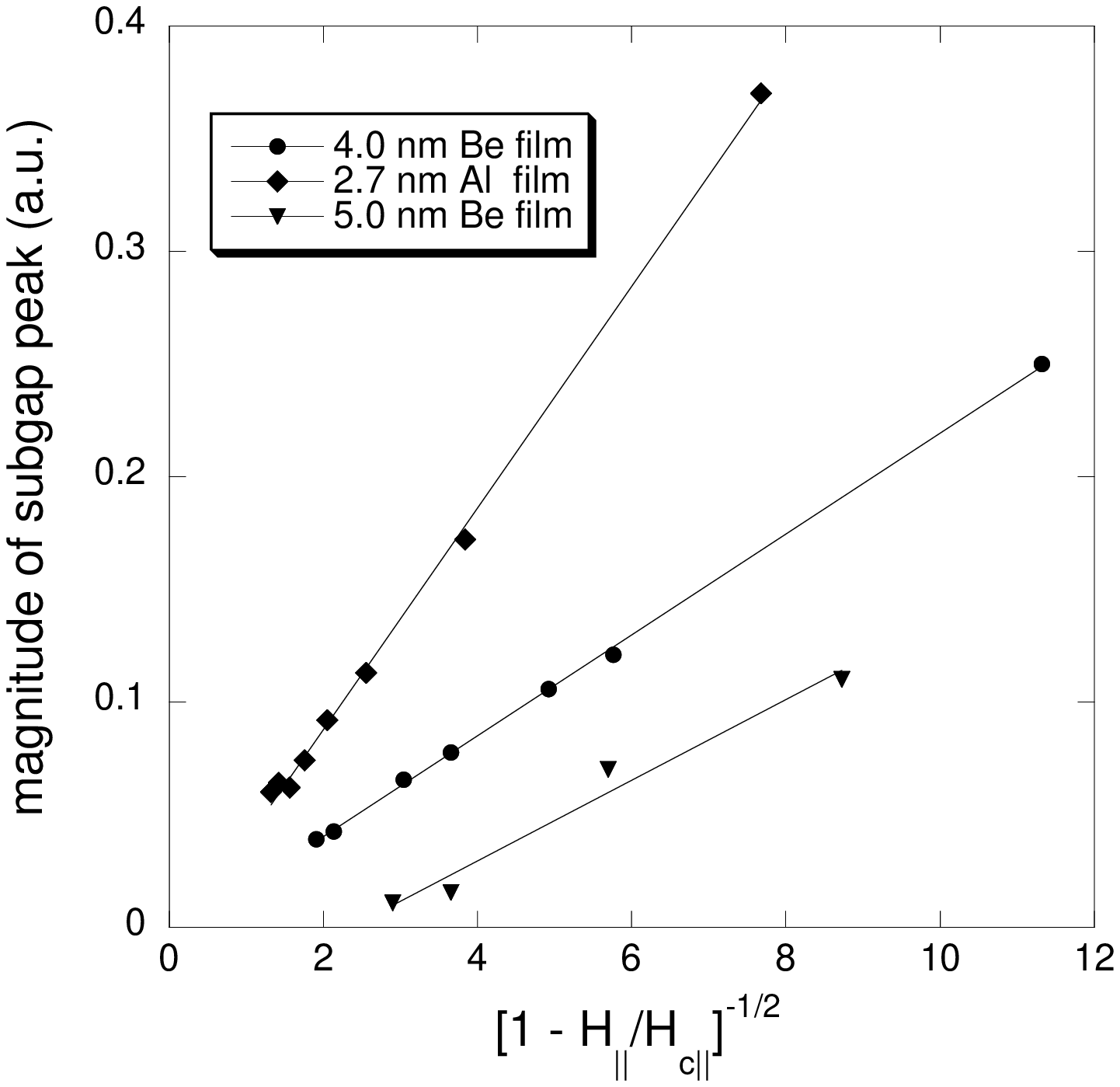}
\caption{\label{fig:epsart}  Scaling behavior of the magnitude of the subgap peak near $H_{c||}$.  The low
temperature normal state sheet resistance of the 4-nm Be film and the 2.7-nm Al film were $\sim1$ k$\Omega$. 
The resistance of the 5-nm Be film was $\sim0.3 $ k$\Omega$.  The parallel critical field values were
$H_{c||}=$ 5.900 T, 1.144 T, and 0.897 T for the Al film, 5-nm Be film, and the 4-nm Be film, respectively.}
\newpage
\end{figure}



\newpage

\begin{references}

\bibitem{UGe2} S. Saxena {\it et al.}, Nature $\bf 406$, 587 (2000).

\bibitem{ZrZn2} C. Pfleiderer {\it et al.}, Nature $\bf 412$, 58 (2001).

\bibitem{FFLO} P. Fulde and R.A. Ferrell, Phys. Rev. $\bf135$, A550 (1964); A.I. Larkin and Y.N. Ovchinnikov, J. Exptl.
Theoret. Phys. (USSR) $\bf47$, 1136 (1964), [Sov. Phys. JETP $\bf20$, 762, (1965)].

\bibitem{CeCoIn} A. Bianchi {\it et al.}, Phys. Rev. Lett. $\bf 89$, 137002 (2002); H.A. Radovan {\it et al.},
Nature {\bf 425}, 51 (2003).

\bibitem{Giaever} I. Giaever, Phys. Rev. Lett. $\bf 5$, 147 (1960).

%

\bibitem{AleinerPRL} I.L. Aleiner and B.L. Altshuler, Phys. Rev. Lett. $\bf 79$, 4242 (1997).

\bibitem{Clogston} A.M. Clogston, Phys. Rev. Lett. $\bf 9$, 266 (1962); B.S. Chandrasekhar, Appl. Phys. Lett. $\bf
1$, 7 (1962).

\bibitem{Fulde} P. Fulde, Adv. Phys. $\bf 22$, 667 (1973).

\bibitem{AdamsAl} W. Wu and P.W. Adams, Phys. Rev. Lett. $\bf 73$, 1412 (1994).

\bibitem{MeserveyMixing} P.M. Tedrow and R. Meservey, Phys. Rev. Lett. $\bf 27$, 919 (1971). R. Meservey, P.M.
Tedrow, and R.C. Bruno, Phys. Rev. B $\bf 11$, 4224 (1975).



\bibitem{MeserveySO} R. Meservey and P.M. Tedrow, Phys. Rev. Lett. $\bf 41$, 805 (1978).

\bibitem{Knight} R.H. Hammond and G.M. Kelly, Phys. Rev. Lett. $\bf 18$, 156 (1966).

\bibitem{AlTCP} V.Yu. Butko and P.W. Adams, Phys. Rev. Lett. $\bf 83$, 3725 (1999).

\bibitem{BeTCP} P.W. Adams, P. Herron, and E.I. Meletis, Phys. Rev. B $\bf 58$, R2952 (1998).

\bibitem{FBTA} V. Yu. Butko, P.W. Adams, and I.L. Aleiner, Phys. Rev. Lett. $\bf 82$, 4284 (1999); P.W. Adams and
V.Yu. Butko, Physica B $\bf 284$, 673 Part 1 (2000); Hae-Young Kee, I.L. Aleiner, and B.L. Altshuler, Phys. Rev. B
$\bf 58$, 5757 (1998).

\bibitem{Inversion} L.P. Gorkov and E.I. Rashba, Phys. Rev. Lett $\bf 87$, 037004 (2001); S.K. Yip, Phys. Rev. B $\bf 65$,
144508 (2002).

\bibitem{Tinkham} M. Tinkam, {\it Introduction to Superconductivity} (McGraw-Hill, New York, 1996).


\bibitem{FFLO2} K. Yang and S.L. Sondhi, Phys. Rev. B $\bf57$, 8566 (1998).




\newpage

\end{references}
\end{document}